# Enabling all-to-circular polarization upconversion by nonlinear chiral metasurfaces with rotational symmetry

Dmitrii Gromyko,[1,2] Jun Siang Loh,[1] Jiangang Feng,[2] Cheng-Wei Qiu,[2, *] and Lin Wu[1,3, *]

[1] *Science, Mathematics, and Technology (SMT), Singapore University of Technology and Design (SUTD), 8 Somapah Road, Singapore 487372.*

[2] *Department of Electrical and Computer Engineering, National University of Singapore, 4 Engineering Drive 3, Singapore 117583*

[3] *Institute of High Performance Computing, A*STAR (Agency for Science, Technology, and Research), 1 Fusionopolis Way, No. 16-16 Connexis, Singapore 138632.*

[chengwei.qiu@nus.edu.sg](chengwei.qiu@nus.edu.sg); [lin_wu@sutd.edu.sg](lin_wu@sutd.edu.sg)

We implement a stacking strategy in designing chiral metasurfaces with high rotational symmetry, enabling quasi-bound-in-the-continuum (quasi-BIC) resonances characterized by absolute chirality. The rotational symmetry allows a circularly polarized pump to be converted into a circularly polarized nonlinear signal. Meanwhile, our bilayered metasurface can be engineered to respond solely to one selected circular polarization. Consequently, integrating resonant chiral response and rotational symmetry endows a unique category of metasurfaces to upconvert any linear or unpolarized pump into a circularly polarized nonlinear signal. Our results reveal that when such a metasurface is subjected to a linearly polarized pump, the intensity ratio of the resultant circularly polarized signals varies with the order of the nonlinear process. Counterintuitively, this ratio scales as the fourth power of the local field enhancement in the second harmonic process and the second power in the third harmonic process. Our work offers a comprehensive theoretical description of the nonlinear processes in chiral structures with rotation and provides universal guidelines for designing nonlinear all-dielectric metasurfaces with a strong chiral response.

**INTRODUCTION**

An object is termed chiral if it is non-superposable on its mirror images[1,2]. Circularly polarized (CP) electromagnetic waves are solutions of Maxwell's equations of absolute chirality: right circularly polarized (RCP) and left circularly polarized (LCP) waves form an orthonormal basis of polarizations while being mirror images of each other. Chirality in optical systems often leads to distinct interactions with CP waves. For instance, circular dichroism CD = $(T^R - T^L)/(T^R + T^L)$, with $T^{R,L}$ the transmission under RCP and LCP pumps is often used as an indicator of chirality. Optical systems and resonators exhibiting chiral response enjoy a vast variety of applications, such as lasing[3], photodetection[4], and sensing[5]. However, chiroptical effects are very weak in natural materials, prompting the development of artificial structures with a pronounced chiral response.

The concept of engineering artificially patterned structures such as photonic crystals, metamaterials, and metasurfaces has reshaped the landscape of devices with tailored optical responses, often unattainable with conventional materials[6,7]. In particular, it has been shown that ultra-thin plasmonic nonlinear metasurfaces with proper rotational symmetry generate CP nonlinear signals when pumped by a CP wave at normal incidence[8]. A plasmonic metasurface with a four-fold rotational symmetry converts an RCP (or LCP) signal into an LCP (or RCP) third harmonic (TH) signal in the transmission geometry[9]. A similar conversion occurs in plasmonic structures with three-fold rotational symmetry, generating a second harmonic (SH) signal[10]. When rotational symmetry is combined with strong chirality, selective enhancement of one CP component can generate a CP nonlinear signal even under a linearly polarized (LP) pump[11]. Although these effects were first demonstrated in plasmonic metasurfaces, an inherent and universal drawback across all plasmonic systems is the considerable absorption losses, which fundamentally constrain device efficiency.

On the other hand, all-dielectric metamaterials appear to be a compelling platform for nonlinear applications due to near-zero material absorption[12-17]. The outstanding feature of all-dielectric devices is the ability to create resonances with ultra-high quality factors (Q-factors) that arise due to the interference of diffracted waves in a periodically patterned medium[18]. A particularly powerful approach to achieve a resonance with a giant tunable Q-factor is to leverage the transition of a non-radiative, bound-in-continuum (BIC) mode with Q = ∞ into a radiative quasi-BIC mode with finite Q-factor[19]. This transition is induced through fine perturbations that alter the structure's symmetry. The characteristics of these resonances are dictated by the perturbation strength and the nature of the symmetry alteration[20]. Theoretically, the Q-factor and local field enhancement can be arbitrarily large given a completely lossless system[21]. All-dielectric metasurfaces possessing quasi-BIC resonances have shown immense potential in achieving giant local field enhancement and, consequently, strong nonlinear signal conversion in ultra-thin metasurfaces[22-24]. Very recently, several metasurface designs with high chirality of resonant modes were proposed[25-32], naturally motivating us to explore the next designs with rotational symmetry and nonlinear response. In principle, all-dielectric metasurfaces with three key components: rotational symmetry, intrinsically chiral high-Q resonances, and nonlinear response should be capable of converting an arbitrarily polarized pump into a purely CP nonlinear signal without the conversion efficiency limit from material losses. Contrarily, the absence of rotational symmetry results in an arbitrary elliptical polarization of the nonlinear signal, while the absence of strong intrinsic chirality does not allow to separate and enhance the preferred circularly polarized component of the pump. To emphasize the exceptionality of such metasurfaces, we summarized the most representative all-dielectric metasurface designs based on the presence of these three components (see FIG. 1a). We envision that such metasurfaces can be used as compact sources of circularly polarized radiation emitting in hard-to-reach wavelength ranges, for example, in the UV range, which is important for a multitude of experiments involving CD[33]. While accessing certain wavelength ranges remains a challenge due to the scarcity of suitable light sources and bulkiness of optical devices such as polarizers, compact nonlinear optical devices appear as a promising solution[14]. Nonlinear metasurfaces capable of all-to-circular upconversion may find applications in chiral sensing[34], CD spectroscopy of molecules[35-37] and nanostructures[38,39].

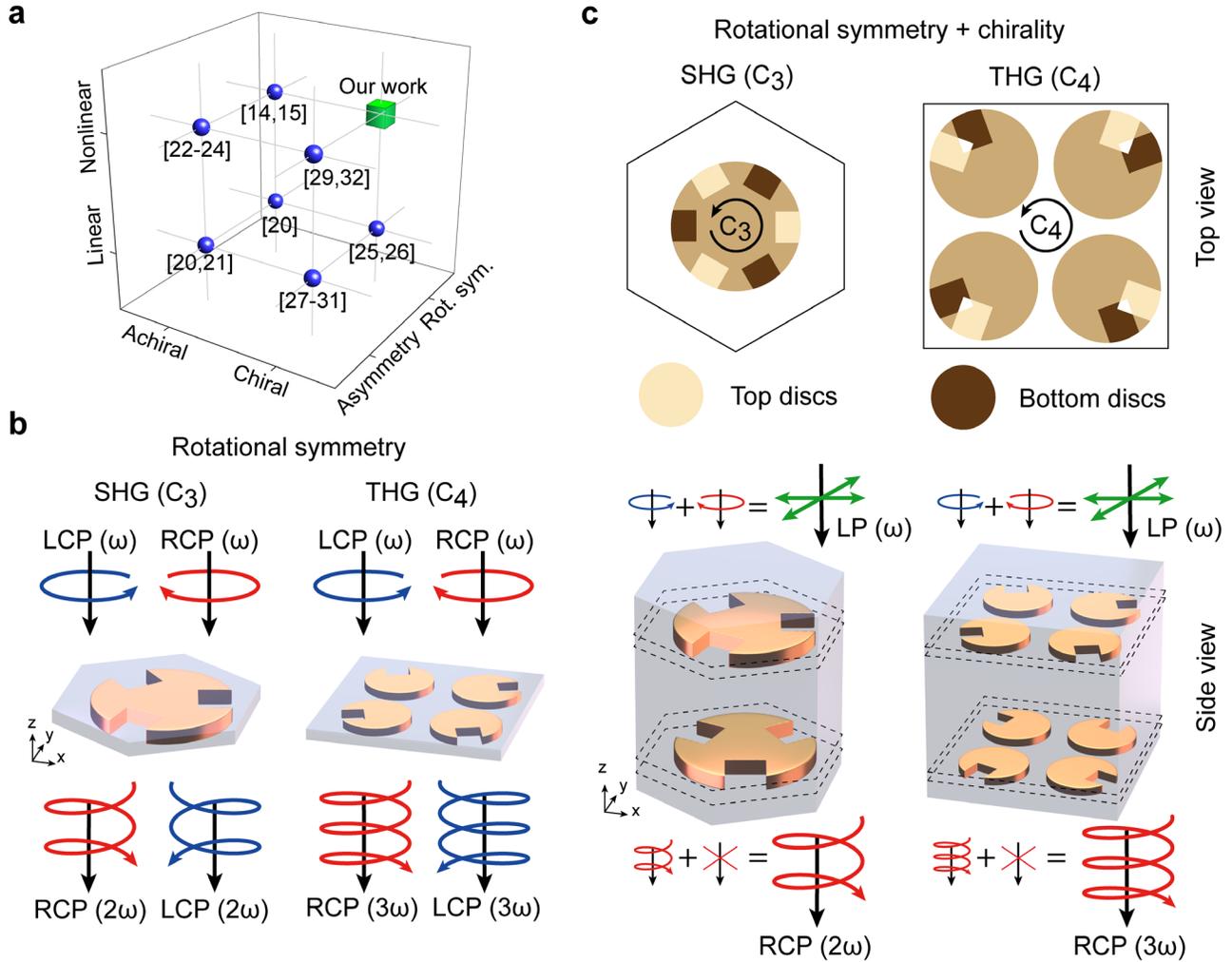

FIG. 1. **Circular upconversion from chiral metasurfaces with rotational symmetry. (a)** Landscape of the all-dielectric metasurface designs by their functionality. Numbers in brackets indicate Ref. numbers. Three key criteria are used: chiral/achiral, absence/presence of rotational symmetry, and absence/presence of nonlinear response. Nonlinear metasurfaces with chirality and rotational symmetry (green cube) are capable of all-to-circular polarization upconversion. **(b)** Generation of CP nonlinear signal from achiral metasurfaces with rotational symmetry under CP pump, where the handedness of the nonlinear signal is opposite to that of the pump: $C_3$ (or $C_4$) rotational symmetry around the vertical axis for SH (or TH) generations. **(c)** Generation of CP nonlinear signal from chiral metasurfaces with rotational symmetry under an LP (composed of RCP and LCP) pump: $C_3$ (or $C_4$) rotational symmetry around the vertical axis for SH (or TH) generations. Only the resonantly enhanced component produces a strong nonlinear signal with circular polarization due to the preserved rotational symmetry.

This work proposes stacked all-dielectric nonlinear metasurfaces with three-fold ($C_3$) or four-fold ($C_4$) rotational symmetry that enjoy quasi-BIC resonances of absolute chirality. We demonstrate that degenerate quasi-BIC resonant modes play a key role in second-harmonic generation (SHG) and third-harmonic generation (THG) within these structures due to their ultra-high Q-factors. We also provide a coupling theory for these modes and derive the conditions of the absolute chirality. Full-wave simulations of the metasurface with optimized chirality demonstrate the conversion of a linearly polarized pump into a nonlinear signal with a near-unity degree of circular polarization. Using symmetry considerations, we reveal that the SH process is preferable to the TH process for such nonlinear polarization conversion as the degree of circular polarization grows faster with the resonance amplitude in the case of the SH process.

**RESULTS**

**GENERALIZED THEORY FOR NONLINEAR SIGNAL POLARIZATION**

The operational principle of the nonlinear metasurfaces with $C_3$ and $C_4$ rotational symmetry is presented in FIG. 1b. If both the material crystalline lattice and the structure have a common vertical axis of symmetry, a normally incident CP plane wave is converted to a nonlinear signal with opposite handedness in the transmission geometry according to the angular momentum conservation law[8,40]. On top of that, if the structure also exhibits a resonant chiral response at pump frequency, waves of the mismatching CP are suppressed and produce negligible nonlinear signal compared to the waves coupled to the resonance (FIG. 1c). One can immediately deduce that with a chiral resonance of a sufficient local field enhancement, the metasurface generates a CP nonlinear signal under an LP or any arbitrarily polarized pump. To provide a rigorous guideline, we first derive a generalized theory linking the amplitude of the chiral resonant mode and polarization of the nonlinear signal.

We assume that both the metasurface and the nonlinear material are $C_3$-symmetric with respect to the vertical z-axis; the metasurface possesses a highly chiral resonance at pump frequency $\omega$ that predominantly couples to LCP waves. Local second-order nonlinear polarization of the material at the doubled frequency is related to the local electric fields **E** induced in the metasurfaces at the pump frequency as $\mathbf{P}^{2\omega} = \varepsilon_0 \hat{\chi}^{(2)} : \mathbf{E}^2$. The concept can be easily generalized to the $C_4$ case using the third-order nonlinear polarization $\mathbf{P}^{3\omega} = \varepsilon_0 \hat{\chi}^{(3)} : \mathbf{E}^3$. Note that in these equations, $\hat{\chi}^{(2),(3)}$ denote the microscopic nonlinear susceptibility tensors of the materials constituting the artificial structures. In this case, an RCP pump induces weak resonant $a_R$ and weak background nonresonant fields with amplitudes $a_R$ and $a_{bg}$, while an LCP wave generates strong resonant fields with amplitude $a_L \gg a_R, a_{bg}$. Due to the preserved rotational symmetry of the structure, the SH signal emitted downwards normally to the metasurface will have an inverted polarization: under the LCP (or RCP) pump, the metasurface generates an RCP (or LCP) SH signal with intensity $I_L^{2\omega}$ (or $I_R^{2\omega}$), where the subscript indicates the pump polarization. At the same time, the intensity of the SH signal is proportional to $|\mathbf{P}^{2\omega}|^2$ or to the 4th power of the field amplitude at the pump frequency:

$$I_L^{2\omega} \propto |a_L|^4, \quad I_R^{2\omega} \propto |a_R + a_{bg}|^4. \tag{1}$$

This leads to the pronounced nonlinear circular dichroism:
$$\mathrm{CD}^{2\omega} = \frac{I_R^{2\omega} - I_L^{2\omega}}{I_R^{2\omega} + I_L^{2\omega}} = -1 + C|a_R + a_{bg}|^4/|a_L|^4. \tag{2}$$

Here and further, $C$ denotes a constant in general that in each equation depends on the specific components of the nonlinear susceptibility tensor. Similarly, for the $C_4$-symmetric structure made of $C_4$-symmetric material, the intensities scale as $|\mathbf{P}^{3\omega}|^2$ or as the 6th power of the field amplitude:
$$I_L^{3\omega} \propto |a_L|^6, \quad I_R^{3\omega} \propto |a_R + a_{bg}|^6, \tag{3}$$

and
$$\mathrm{CD}^{3\omega} = \frac{I_R^{3\omega} - I_L^{3\omega}}{I_R^{3\omega} + I_L^{3\omega}} = -1 + C|a_R + a_{bg}|^6/|a_L|^6. \tag{4}$$

When the structure is illuminated by an LP wave (FIG. 1c), the local electric fields inside the metasurface at the pump frequency can be presented as a sum of an RCP- and LCP-generated fields $\mathbf{E}(\mathbf{r}) = \mathbf{E}_R(\mathbf{r}) + \mathbf{E}_L(\mathbf{r})$. One can write the three terms of the induced nonlinear polarization as:
$$\mathbf{P}^{2\omega} = \varepsilon_0 \hat{\chi}^{(2)} : (\mathbf{E}_R^2 + 2\mathbf{E}_R \mathbf{E}_L + \mathbf{E}_L^2). \tag{5}$$

This polarization can be treated as a source of nonlinear signals. According to the Lorentz reciprocity theorem[41], the intensity of the RCP and LCP signal emitted by a periodic dipole density $\mathbf{P}^\Omega(\mathbf{r})$ to the top or bottom half-space is proportional to:
$$I_{R,L}^{\Omega(t,b)} \propto \left| \int \mathbf{E}_{R,L}^{\Omega(t,b)}(\mathbf{r}) \cdot \mathbf{P}^\Omega(\mathbf{r}) d^3\mathbf{r} \right|^2, \tag{6}$$

where $\mathbf{E}_{R,L}^{\Omega(t,b)}(\mathbf{r})$ is the electric field induced in the structure by a hypothetical incident RCP or LCP plane wave at the frequency of the dipole's oscillation $\Omega$. The integral is evaluated over the unit cell. Superscripts $t$ and $b$ indicate that the waves come from the top or bottom.

At this point, we introduce an operator $R_\phi$ of clockwise rotation by angle $\phi$ around the $z$-axis (top view). Due to $C_3$ symmetry, the structure and the nonlinear susceptibility tensor are invariant under $R_{2\pi/3}$ rotation while the RCP- and LCP-induced fields in the structure transform like the electric fields of the RCP and LCP plane waves, *i.e.*,
$$R_{2\pi/3}\left(\mathbf{E}_{R,L}^{(t)}\right) = \exp(\pm i 2\pi/3)\mathbf{E}_{R,L}^{(t)}, \quad R_{2\pi/3}\left(\mathbf{E}_{R,L}^{(b)}\right) = \exp(\mp i 2\pi/3)\mathbf{E}_{R,L}^{(b)}. \tag{7}$$

Accordingly, the three terms of nonlinear polarization in Eq. *(5)* generated by an LP plane wave incident from the top under $R_{2\pi/3}$ rotation acquire phase factors $\exp(i 4\pi/3)$, $\exp(0)$, and $\exp(-i 4\pi/3)$, respectively. To obtain the intensities of LCP and RCP components of the nonlinear signal in the transmission geometry, we consider local fields $\mathbf{E}_{L,R}^{2\omega(b)}(\mathbf{r})$ excited from the bottom by an LCP and RCP plane waves with frequency $2\omega$. We also assume that fields $\mathbf{E}_{L,R}^{2\omega(b)}(\mathbf{r})$ to be of comparable amplitudes, which means the absence of strong chiral resonances at the doubled frequency. Substituting these fields into the reciprocity theorem, we derive (superscript b is omitted for brevity):

$$I^{2\omega}_{\text{LP(L)}} \propto \left| \int \mathbf{E}_{\text{L}}^{2\omega(b)} \cdot \hat{\chi}^{(2)} : \mathbf{E}_{\text{R}}^2 d^3\mathbf{r} \right|^2, \tag{8}$$

$$I^{2\omega}_{\text{LP(R)}} \propto \left| \int \mathbf{E}_{\text{R}}^{2\omega(b)} \cdot \hat{\chi}^{(2)} : \mathbf{E}_{\text{L}}^2 d^3\mathbf{r} \right|^2. \tag{9}$$

The nonlinear signals associated with the cross-polarization terms destructively interfere so that the ratio of the SH signal CP intensities scales as the fourth power of the field enhancement:

$$I^{2\omega}_{\text{LP(R)}}/I^{2\omega}_{\text{LP(L)}} \propto |a_{\text{L}}|^4/|a_{\text{R}} + a_{\text{bg}}|^4. \tag{10}$$

As a result, an LP pump generates an SH signal with a **degree of circular polarization (DCP)**:

$$\text{DCP}^{2\omega(b)} = \frac{I^{2\omega}_{\text{LP(R)}} - I^{2\omega}_{\text{LP(L)}}}{I^{2\omega}_{\text{LP(R)}} + I^{2\omega}_{\text{LP(R)}}} = 1 - C|a_{\text{R}} + a_{\text{bg}}|^4/|a_L|^4 \approx -\text{CD}^{2\omega} \tag{11}$$

One can estimate that even a modest enhancement of the LCP signal $|a_{\text{L}}|/|a_{\text{R}} + a_{\text{bg}}| = 5$ ensures near-unity DCP>0.99 of SH signal (assuming $C = 2$).

On the other hand, in the case of $C_4$-structure, the nonlinear polarization produced by the third-order nonlinear susceptibility $\hat{\chi}^{(3)}$ can be expanded as:

$$\mathbf{P}^{3\omega}(\mathbf{r}) = \varepsilon_0 \hat{\chi}^{(3)} : (\mathbf{E}_{\text{R}}^3 + 3\mathbf{E}_{\text{R}}^2 \mathbf{E}_{\text{L}} + 3\mathbf{E}_{\text{R}} \mathbf{E}_{\text{L}}^2 + \mathbf{E}_{\text{L}}^3). \tag{12}$$

Similarly to the $C_3$ case, the electric fields in the $C_4$ structure acquire phase factor $\exp(\pm i\pi/2)$ under the rotation operator $R_{\pi/2}$, thereby the reciprocity theorem gives:

$$\begin{aligned} I^{3\omega}_{\text{LP(L)}} &\propto \left| \int \mathbf{E}_{\text{L}}^{3\omega(b)} \cdot (\hat{\chi}: \mathbf{E}_{\text{R}}^3 + 3\hat{\chi}^{(3)} : \mathbf{E}_{\text{R}} \mathbf{E}_{\text{L}}^2) d^3\mathbf{r} \right|^2, \\ I^{3\omega}_{\text{LP(R)}} &\propto \left| \int \mathbf{E}_{\text{R}}^{3\omega(b)} \cdot (3\hat{\chi}: \mathbf{E}_{\text{R}}^2 \mathbf{E}_{\text{L}} + \hat{\chi}^{(3)} : \mathbf{E}_{\text{L}}^3) d^3\mathbf{r} \right|^2, \end{aligned} \tag{13}$$

and all other integrals cancel out due to symmetry considerations. The ratio of the RCP and LCP TH intensities produced by a linearly polarized pump scales as the second power of the field enhancement:

$$I^{3\omega}_{\text{LP(R)}}/I^{3\omega}_{\text{LP(L)}} \propto |a_{\text{L}}|^2/|a_{\text{R}} + a_{\text{bg}}|^2. \tag{14}$$

Comparing Eqs. *(14)* to (10), we can expect the second-order nonlinearity supported by a chiral resonance to be advantageous to the third-order nonlinearity in terms of the nonlinear signal DCP; as for the TH signal, we obtain:

$$\text{DCP}^{3\omega(b)} \approx 1 - C|a_{\text{R}} + a_{\text{bg}}|^2/|a_{\text{L}}|^2 \neq -\text{CD}^{3\omega}. \tag{15}$$

Eqs. *(11)* and *(15)* show that the SH signal DCP tends to unity much faster than that for TH signal. For example, one can estimate DCP=0.92 for the TH signal using the same relative enhancement $|a_{\text{L}}|/|a_{\text{R}} + a_{\text{bg}}| = 5$ and $C = 2$.

Note that in the reflection geometry, the polarization of the nonlinear signal should be the opposite of that in the transmission geometry. Thus, an LP pump generates an RCP nonlinear signal propagating upwards. Theoretically, we confirm that the proper rotational symmetry of the structure and absolute chirality of a high-Q resonance allow for generating a fully polarized nonlinear signal under a linearly polarized pump. The provided derivations are general and valid for lossless dielectric and lossy plasmonic metasurfaces. Application of the theory to the case of an arbitrary elliptically polarized pump is straightforward: the output nonlinear signal should exhibit high DCP as long as the pump signal has a resonance-matching circular polarization component of sufficient magnitude.

To validate our derivation, we will commence an in-depth numerical analysis using an exemplified $C_4$-symmetric chiral structure for THG. The chosen system's modes serve as an illustrative and analytically accessible starting point. Subsequently, we will extend this analysis to a $C_3$-symmetric system. Finally, we will calculate the dependencies of SH and TH DCPs on material absorption for our optimized metasurface designs and verify the scaling laws in Eqs. (10) and *(14)* derived here.

## $C_4$-SYMMETRIC CHIRAL METASURFACES FOR THIRD HARMONIC GENERATION

The engineering of a chiral resonance is a challenging and self-sufficient problem that can be perplexed by the requirement of rotational symmetry. To create a quasi-BIC resonance with absolute chirality, we use stacked all-dielectric metasurfaces. Due to the vertical separation, stacked geometries introduce a natural phase shift into the electromagnetic wave scattering problem, while the rotational symmetry of both top and bottom metasurfaces results in the rotational symmetry of the stacked structure. Moreover, the ability to control the resonant properties of the constituent metasurfaces is a key to controlling the resonant properties of the whole structure.

The model structure consists of the top and bottom flat metasurfaces with $C_4$ symmetry (FIG. 1c), where each constituent single-layer metasurface is a two-dimensional (2D) array of discs arranged in a square lattice (Fig. 2a). The centers of the top and bottom discs are aligned, so that the upper discs are directly positioned over the bottom discs. One notch is carved in each disc in such a way as to double the period of the square lattice while maintaining the 4-fold rotational symmetry of the stacked metasurface. The period doubling allows for creating a doubly degenerate radiative mode at Γ-point from a guided mode on the edge of the Brillouin zone. It is worth noting that all non-degenerate modes in $C_4$-symmetric structures are BIC at Γ-point below the frequency of diffraction cut-off and cannot be used[20]. The transition of the optically inactive guided mode into a quasi-guided mode allows one to tune the radiation losses of the mode by changing the dimensions of the notch in the same way as for quasi-BIC resonance. In the following simulations, the top and bottom metasurfaces are characterized by parameters: period $p = 860$ nm, disc radius $R = 190$ nm, disc thickness $h = 150$ nm, notch width $w = 110$ nm and depth $d = 100$ nm, while the **interlayer distance $H$** and **twist angle $\theta$** are selected as tuning parameters for achieving chirality. The discs have a refractive index $n_{\text{disc}} = 3.4$, extinction coefficient $k = 0.003$, and are fully embedded in medium with refractive index $n = 1.45$. The nonlinear polarization of the discs is defined as $\mathbf{P}^{3\omega} = \varepsilon_0 \chi^{(3)} |\mathbf{E}|^2 \mathbf{E}$, with $\chi^{(3)} = 2.45 \cdot 10^{-19}$ m²/V² which is close to that of α-Si[17].

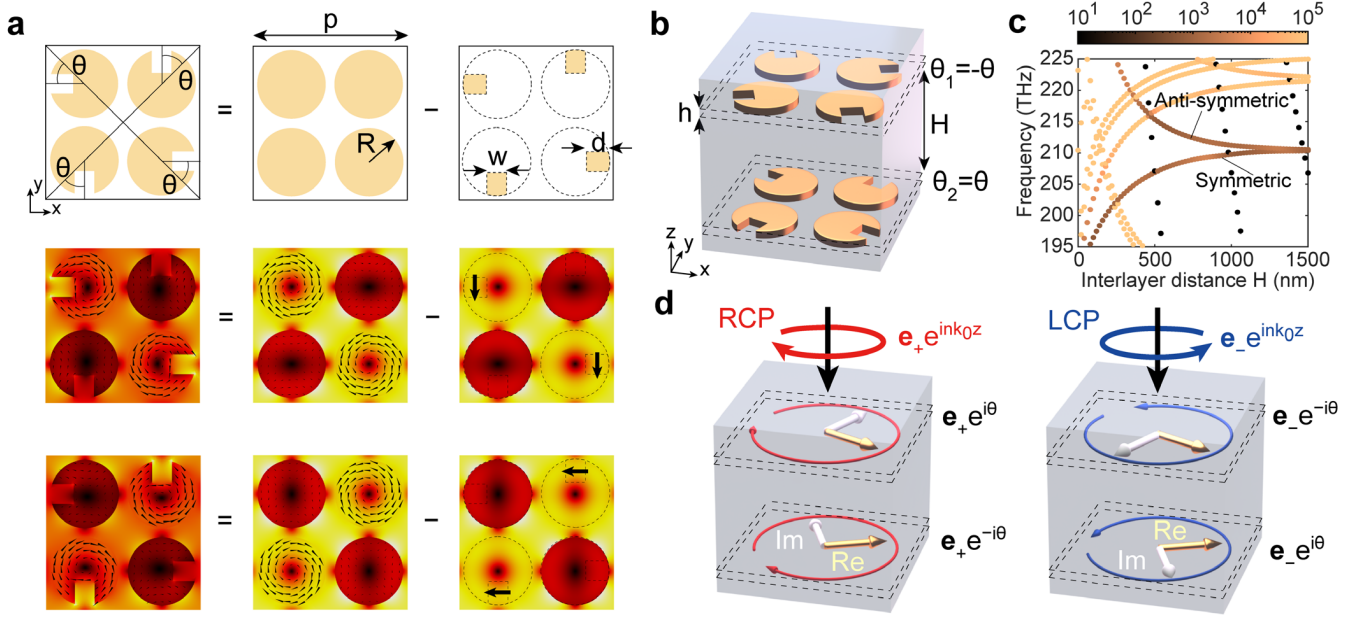

Fig. 2. $C_4$-symmetric chiral structures for third harmonic generation. (a) Schematic illustration of a single-layer metasurface (unit cell) with $C_4$ symmetry and the electric fields of the degenerate quasi-BIC mode found in a single-layer metasurface. The resonant mode can be effectively decomposed into the unperturbed field of the BIC mode and the perturbation field. (b) A unit cell of the stacked $C_4$ metasurface possessing a chiral quasi-BIC resonance. Notches in the top and bottom metasurfaces are cut at angles $\mp\theta$. (c) Dependence of the resonant modes' energies on the separation distance. The color indicates the radiative Q-factor of the modes. Two modes with $Q \approx 10^3$ are the quasi-BIC degenerate modes that evolve from symmetric and anti-symmetric BIC modes of the structure without notches. (d) Schematic illustration of the circular dipoles representing the RCP-induced and LCP-induced resonant fields of the symmetric mode in the top and bottom metasurfaces.

The resonant properties of the stacked structure can be efficiently described using the resonant dipoles model (see the details in SI-I). In this model, the electric field $\mathbf{E}_{res}$ of a resonant mode $\mathbf{E}_{res}$ is integrated over the volume of the top or bottom discs $V_{t,b}$ and later are treated as dipoles:

$$\mathbf{D}_{t,b} = \int_{V_{t,b}} \mathbf{E}_{res}(\mathbf{r})d^3\mathbf{r}. \tag{16}$$

When the structure is excited by a normally incident plane wave $\mathbf{E}_{inc}e^{ink_0 z + i\omega t}$ (here, $k_0$ is the wavenumber of the pump in the free space and $n$ in the refractive index of the embedding medium), the amplitude of the resonant mode is proportional to:

$$a \propto \mathbf{D}_t \cdot \mathbf{E}_{inc} + \mathbf{D}_b \cdot \mathbf{E}_{inc} e^{-ink_0(H+h)}, \tag{17}$$

where the phase factor $e^{-ink_0(H+h)}$ accounts for the propagation of the incident wave between the centers of the top and bottom metasurfaces. Let us consider an RCP and LCP waves incident on a metasurface hosting a chiral resonance. In what follows, we use the following CP basis vectors:

$$\mathbf{e}_\pm = \{1, \pm i\}/\sqrt{2}. \tag{18}$$

For the plane waves coming from the top (bottom), *i.e.*, negative (positive) *z*-direction of propagation, $\mathbf{e}_\pm$ represents the RCP (LCP) and LCP (RCP) plane waves. We define the **resonance chirality** (**RC**) value using the resonance amplitudes $a_{R,L}$ under RCP and LCP pump as:

$$\text{RC} = \frac{|a_R|^2 - |a_L|^2}{|a_R|^2 + |a_L|^2}, \tag{19}$$

so that the resonances of absolute chirality $\text{RC} = \pm 1$ are fully decoupled from top LCP or RCP waves, respectively.

To explicitly derive the conditions of absolute chirality, we leverage the resonant dipole coupling model. Since we are dealing with quasi-BIC modes, their coupling to the far field is solely defined by the size and position of the notches that break the initial symmetry. We first consider the modes of the single-layer metasurfaces and later use them to construct the modes in the stacked structure. The fields of the degenerate quasi-BIC mode with frequency 210 THz of a single-layer metasurface are shown in Fig. 2a (1st column). These fields can be decomposed into a sum of the unperturbed fields of the BIC mode (2nd column) and the perturbation field (3rd column). Due to the symmetrical distribution of the electric fields inside the discs, the resultant effective dipoles for the two modes can be approximated with perpendicular dipoles whose direction is controlled by the notch angle $\theta$:

$$\mathbf{D}_1 = \{\cos(\theta - \pi/4), \sin(\theta - \pi/4)\}, \quad \mathbf{D}_2 = \{\cos(\theta + \pi/4), \sin(\theta + \pi/4)\}. \tag{20}$$

It is also straightforward to construct their CP combinations that are excited by the RCP and LCP pump from the top:

$$\mathbf{D}_+ = \mathbf{D}_-^* = (\mathbf{D}_1 + i\mathbf{D}_2)/\sqrt{2} = e^{-i(\theta - \pi/4)} \mathbf{e}_+. \tag{21}$$

The structure of stacked metasurfaces without notches possesses a horizontal plane of reflection symmetry; thus, the quasi-BIC modes evolve from the reflection-symmetric and anti-symmetric degenerate BIC modes. The notches of the top and bottom metasurface are cut at angles $\mp\theta$ with respect to the diagonals of the unit cell (Fig. 2b). Such structure perturbation breaks the out-of-plane mirror symmetry and unlocks the possibility of controlling the resonance chirality. The typical dependence of the hybridized modes frequencies on the separation distance is shown in Fig. 2c. Two hybridized modes with the $Q \approx 10^3$ converge to the frequency of the quasi-BIC single-layer mode in the limit $H > 1000$ nm. We may also notice several modes with $Q < 10$ with a steep dependence of the resonant energy on interlayer distance. These are the Fabry-Perot modes of the stacked structure formed by the weak background reflection of the top and bottom metasurfaces. As we will be shown further, the chirality of the hybridized quasi-BIC modes is possible when the frequencies of these modes coincide with those of the Fabry-Perot modes.

We choose the symmetric hybridized mode as an example and consider it a superposition of the modes of the top and bottom single-layer metasurfaces. Since the notches at the top and bottom metasurfaces are cut at angles $\mp\theta$, the resonant dipoles of the top and bottom metasurfaces (Fig. 2d) are given as:

$$\begin{aligned}
\mathbf{D}_{t,+}^{(0)} &= e^{i\theta}\mathbf{e}_+, \quad \mathbf{D}_{b,+}^{(0)} = e^{-i\theta}\mathbf{e}_+ \\
\mathbf{D}_{t,-}^{(0)} &= e^{-i\theta}\mathbf{e}_-, \quad \mathbf{D}_{b,-}^{(0)} = e^{i\theta}\mathbf{e}_-
\end{aligned} \tag{22}$$

up to a common factor. Considering the coupling of the modes, the total resonant fields in each single-layer metasurface are written as a sum of the fields in Eq. *(22)* and the coupling-induced fields. The additional fields induced in each metasurface due to the far-field coupling are proportional to the fields of the counterpart metasurface with a factor $r\exp(-ink_0H)$, where $r$ is a background reflection coefficient of a single-layer metasurface (see the details in SI-II). In short, for the hybridized modes of the stacked structure, the resonant amplitudes in the top and bottom metasurfaces can be written as:

$$\begin{aligned}
\mathbf{D}_{t,\pm} &= \mathbf{D}_{t,\pm}^{(0)} + \mathbf{D}_{b,\pm}^{(0)} r e^{-ink_0H}, \\
\mathbf{D}_{b,\pm} &= \mathbf{D}_{b,\pm}^{(0)} + \mathbf{D}_{t,\pm}^{(0)} r e^{-ink_0H}.
\end{aligned} \tag{23}$$

Substituting CP resonant dipoles into Eq. *(17)* (see details in SI-I) and using the notation $r = |r|e^{i\psi}$, we find that either $a_L = 0$ or $a_R = 0$ is achieved when:

$$e^{\pm 2i\theta} = -\frac{1 + |r|e^{-ink_0(2H+h)+i\psi}}{e^{-ink_0(H+h)}(1 + |r|e^{ink_0h+i\psi})}. \tag{24}$$

This relation is fulfilled if the absolute value of the right-hand side expression equals one. Thus, nontrivial interlayer distance $H$ and twist angle $\theta$ that ensure the absolute chirality should satisfy the following equations:

$$\begin{aligned}
nk_0H - \psi &= \pi m, \; m \in \mathbb{Z} \Leftrightarrow re^{-ink_0H} = \pm|r|; \\
nk_0(H+h) \pm 2\theta &= \pi + 2\pi m' + 2\cdot\arg(1 + |r|e^{ink_0h+i\psi}).
\end{aligned} \tag{25}$$

The first condition on the interlayer distance $H$ suggests that the absolute chirality exists in structures tuned to the full background transmission of the Fabry-Perot type. Based on the energy dependences of the Fabry-Perot modes shown in Fig. 2c, we can expect full chirality near $H$ = 500 nm and $H$ = 1000 nm.

Using COMSOL eigensolver, we simulate the electric fields of the symmetric resonant modes in the stacked structure, calculate the resonant dipoles according to Eq. *(16)* and find the amplitudes of the resonant modes $a_{R,L}$. The RC values for the symmetric hybrid mode in metasurfaces with varied twist angle $\theta$ and interlayer distance $H$ are presented in FIG. 3a. To check the validity of our theory, we substitute the coupled resonant dipoles in Eq. *(23)* into Eq. *(17)* and plot the theoretical RC values in the lower panel of FIG. 3a. We find a satisfactory correspondence between the simulations and the theory and highlight three optimized designs with maximized chirality using symbols. A similar analysis is performed for the anti-symmetric mode in SI-III. From the RC map in FIG. 3a, we select a representative optimized metasurface design with parameters $H$ = 500 nm, $\theta$ = 62$^o$ (indicated by the circular symbol) that exhibits an RCP-decoupled resonance and calculates the linear and nonlinear spectra of the structure. FIG. 3b shows that RCP waves are nonresonantly transmitted with near-unity amplitude, while LCP waves are resonantly absorbed. The linear CD reaches −1 at the resonance as material absorption of the dielectric discs is chosen to be $k$ = 0.003 to embody the critical coupling conditions[26,42], *i.e.*, equality of radiative and nonradiative losses of the resonance.

Next, we calculate the TH signal generated by the CP pump. At the frequency of TH, signal radiation propagates along the normal direction and leaks into the diffraction channels, which are now open. Fortunately, the radiation into different diffraction channels can be easily separated. In what follows, we only focus on the properties of the nonlinear signal in the main diffraction channel. Following the resonance chirality, the RCP pump generates a very weak LCP TH signal, while the LCP pump generates a strong RCP TH signal supported by the chiral quasi-BIC resonant mode (FIG. 3c). Nonlinear CD exhibits close to unity values in a larger range of frequencies in the vicinity of the resonance compared to the linear CD. Finally, when the structure is illuminated by an LP plane wave, the DCP of the nonlinear signal gets close to unity in the vicinity of the resonance, which manifests the generation of an RCP TH signal (FIG. 3d). The DCP curve is highly asymmetric and drops quickly at off-resonant conditions compared to the nonlinear CD. We suppose that such behavior is caused by the complex interference of the nonlinear polarization terms in Eq. *(13)* that contribute to the RCP and LCP components of the TH signal. Other optimized $C_4$-symmetric stacked metasurface designs indicated by triangular and square symbols are analyzed in SI-III. We estimate that the nonlinear signal generated by our nonlinear metasurface is 4 orders higher than the signal generated by an unpatterned nonlinear film and 40 times higher than the signal from a metasurface supporting Mie modes (see SI-IV). Moreover, the nonlinear signal amplification can theoretically achieve higher values in metasurfaces with smaller asymmetry parameters, such as smaller notches in our case, provided that nonradiative losses are negligible[22]. However, in practical scenarios, resonant enhancement is fundamentally limited by nonradiative losses due to material absorption, random scattering from irregularities and fabrication defects, and edge leakage in finite-size samples[23]. Therefore, our simulations are confined to metasurfaces with resonances exhibiting reasonable Q-factors tuned to critical coupling conditions. Additionally, in SI-IV we compare the functionality of stacked metasurfaces with rotational symmetry and reveal that contrarily to quasi-BIC resonances, Mie resonances do not provide absolute chirality of hybridized modes due to limited far-field radiation control.

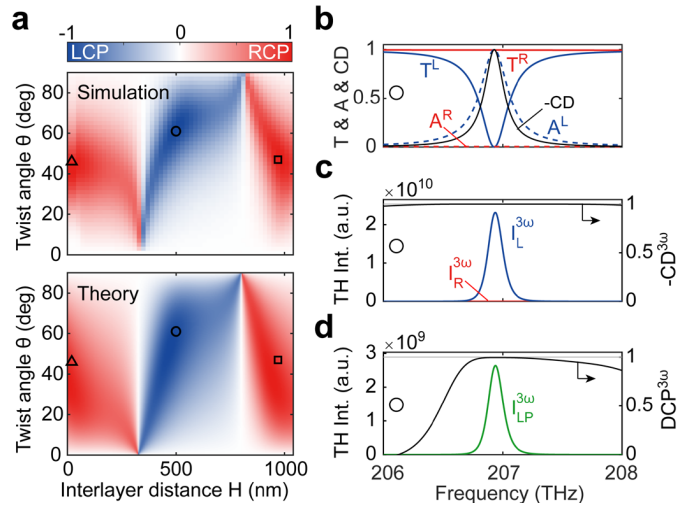

FIG. 3. **Designing $C_4$-symmetric stacked structures for maximized chirality. (a)** Resonance chirality (RC) calculated based on COMSOL eigenmode simulation and analytically using the coupled dipoles model in Eq. (22) and Eq. (23). The black symbols highlight the optimized designs with maximized chirality. **(b)-(d)**

Analysis of one selected design with $H$ = 500 nm, $\theta$ = 62º (circle) in the vicinity of the symmetric quasi-BIC mode with optimized chirality. The extinction coefficient of the dielectric discs is chosen $k$ = 0.003 to ensure the critical coupling conditions and maximize the linear CD. **(b)** Transmission and absorption spectra (solid and dashed lines) under RCP (red) or LCP (blue) pump and linear CD (black line). **(c)** Intensity of the TH signal under RCP (red) or LCP pump (blue) and the nonlinear CD$^{3\omega}$ (black line). **(d)** Intensity of the TH signal under the LP pump (green line) and the DCP$^{3\omega}$ of the TH signal (black line).

## $C_3$-SYMMETRIC CHIRAL METASURFACES FOR SECOND HARMONIC GENERATION

Applying the same stacking approach, we propose a model $C_3$-symmetric chiral structure (FIG. 1c) that consists of the top and bottom flat metasurfaces, where each constituent metasurface is a 2D hexagonal array of high-index dielectric discs possessing 6-fold rotational symmetry. Three notches are cut at the sides of each disc, reducing the symmetry of a single metasurface from $C_{6v}$ to $C_3$ (FIG. 4a). As known, this symmetry breaking engenders a doubly degenerate quasi-BIC mode[20]. While for the $C_{6v}$-symmetric metasurface all modes except E$_1$ mode are BIC at Γ-point below the first diffraction opening, in a $C_3$-symmetric case both degenerate modes E$_{1,2}$ transform into radiative modes E. The electric fields of the BIC mode E$_2$ of the unperturbed structure are depicted in FIG. 4b. Leveraging the transition E$_2$ (BIC) into E (quasi-BIC), we create a resonance with tunable radiative losses while maintaining the three-fold rotational symmetry of the metasurface. In the stacked metasurface, the top discs are positioned exactly above the bottom discs to preserve vertical alignment and rotational symmetry. As in the $C_4$ case, coupling of the degenerate modes of top and bottom metasurfaces produces two degenerate hybridized modes: one mode decays symmetrically along the vertical axis, and another decays asymmetrically. To break the out-of-plane symmetry and achieve a chiral response, the notches of the top and bottom metasurfaces are cut at angles $\pm\theta$ (see FIG. 4c). $C_3$ metasurface has fixed period $p = 1000$ nm, disc radius $R = 300$ nm, disc thickness $h = 80$ nm, notch width $w = 150$ nm, and notch depth $d = 100$ nm. The discs have a model refractive index $n_{\text{disc}} = 4$, extinction coefficient $k = 0.006$, and are fully embedded in medium with refractive index $n$ = 1.45. The nonlinear polarization of the discs is defined as $P_y = -2\varepsilon_0\chi^{(2)}E_yE_x$, $P_y = \varepsilon_0\chi^{(2)}(E_y^2 - E_x^2)$, $P_z = 0$, where we chose a model susceptibility $\chi^{(2)} = 1.6 \cdot 10^{-11}$ m/V. The resonant frequencies of these hybrid modes are provided in SI-V.

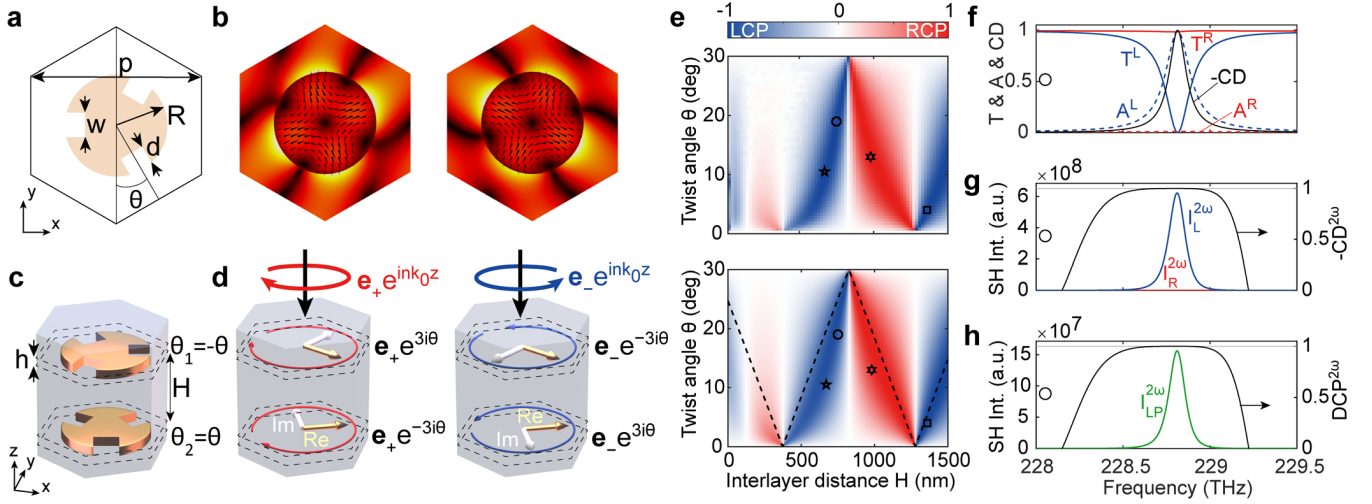

FIG. 4. $C_3$-symmetric chiral structures for second harmonic generation. (a) Schematic illustration of the single-layer metasurface with $C_3$ symmetry (unit cell). (b) The electric fields of the degenerate BIC mode in a single-layer metasurface without notches. (c) A unit cell of the stacked $C_3$ metasurface possessing a chiral quasi-BIC resonance. (d) Schematic illustration of the circular dipoles representing the RCP-induced and LCP-induced resonant fields of the symmetric hybridized mode in the top and bottom metasurfaces. (e) Resonance chirality (RC) calculated based on COMSOL eigenmode simulation and analytically using the coupled dipoles model in Eq. (26) and Eq. (28). The black symbols highlight the optimized designs with maximized chirality. The black dashed line denotes solutions of Eq. (27). (f)-(h) Analysis on one selected $C_3$ design with $H = 750$ nm, $\theta = 19°$ (circle) possessing the resonance of absolute chirality. The extinction coefficient of the dielectric discs is chosen $k = 0.006$ to ensure the critical coupling conditions and maximize the linear CD. (f) Transmission and absorption spectra (solid and dashed lines) under RCP (red) or LCP (blue) pump and linear CD (black line). (g) Intensity of the SH signal under RCP (red) or LCP pump (blue) and the nonlinear $CD^{2\omega}$ (black line). (h) Intensity of the SH signal under the LP pump (green line) and the $DCP^{2\omega}$ of the SH signal (black line).

For the symmetrically decaying mode, the polarization of the RCP-induced and LCP-induced resonant fields in the top and bottom metasurfaces is similar to that in the $C_4$ case (see a schematic illustration in FIG. 4d):

$$\mathbf{D}_{t,+}^{(0)} = e^{3i\theta}\mathbf{e}_+, \quad \mathbf{D}_{b,+}^{(0)} = e^{-3i\theta}\mathbf{e}_+;$$
$$\mathbf{D}_{t,-}^{(0)} = e^{-3i\theta}\mathbf{e}_-, \quad \mathbf{D}_{b,-}^{(0)} = e^{3i\theta}\mathbf{e}_-. \tag{26}$$

A single-layer metasurface, in this case, is chosen such that the background reflection coefficient is negligibly small; thus, the resonant dipoles in the form (26) can be directly used to derive the absolute chirality conditions:

$$nk_0(H + h) \pm 6\theta = \pi + 2m\pi, \quad m \in \mathbb{Z}. \tag{27}$$

Nevertheless, the coupling model can be improved using the near-field-induced resonant fields whose amplitudes decay with distance according to the simple exponential law:

$$\begin{aligned} \mathbf{D}_{t,\pm} &= \mathbf{D}_{t,\pm}^{(0)} + \mathbf{D}_{b,\pm}^{(0)} C e^{-\kappa H}, \\ \mathbf{D}_{b,\pm} &= \mathbf{D}_{b,\pm}^{(0)} + \mathbf{D}_{t,\pm}^{(0)} C e^{-\kappa H}. \end{aligned} \quad (28)$$

Here, $C$ is a fitting parameter of the near-field coupling. The decay constant $\kappa$ is the modulus of the wavevector's $z$-component in the first diffraction channel of the hexagonal lattice:

$$\kappa = \left|k_z^{(1)}\right| = \sqrt{\left(\frac{4\pi}{p\sqrt{3}}\right)^2 - \left(\frac{n\omega}{c}\right)^2}. \quad (29)$$

We demonstrate the COMSOL-simulated and theoretically calculated resonance chirality maps (FIG. 4e) for the symmetric mode in $C_3$ stacks with varied interlayer distance $H$ and twist angle $\theta$. The theoretical values are calculated based on the model in Eq. *(28)* and the definition in Eq. *(17)*. We can observe that optimal twist angles $\theta$ corresponding to the absolute chirality only slightly deviate from those predicted in the approximation of uncoupled modes in Eq. *(27)* (denoted with the dashed line). This is due to the near-zero background reflection coefficient $r$ of a single-layer metasurface. This is also why absolute chirality can be achieved for almost any twist angle that breaks the horizontal mirror symmetry. A similar chirality map for the anti-symmetric mode is provided in SI-V. Interestingly, due to the broad range of structure parameters that provide chirality of the two modes, one can identify the structures that exhibit absolute chirality simultaneously (see SI-V).

We select a $C_3$ metasurface design with parameters $H$ = 750 nm, $\theta$ = 19° (indicated by the circular symbol) to simulate the linear and nonlinear optical spectra under RCP or LCP pump in FIG. 4e (see results for other designs in SI-V). Following the dipole model, the resonance selectively couples to LCP waves (FIG. 4f). As in the $C_4$ case, the range of frequencies where the near-unity nonlinear CD (FIG. 4g) is achieved is wider than that for linear CD (FIG. 4f). The main difference from the $C_4$ case can be found when the structure is illuminated by an LP light (FIG. 4h). We notice that in this case the intensity of the SH signal ($I_{LP}^{2\omega}$ in FIG. 4h) is exactly 4 times smaller than that produced under the LCP pump ($I_L^{2\omega}$ in FIG. 4g). The most outstanding feature is that the nonlinear signal DCP (FIG. 4h) fully coincides with the nonlinear CD (FIG. 4g) in the vicinity of the resonance as predicted by Eq. *(11)*.

**SCALING LAWS OF NONLINEAR SIGNAL INTENSITY**

Finally, we corroborate that the intensity of the nonlinear signal and its RCP and LCP components scale as predicted by Eqs. (10), *(14)* that is, the ratio of the RCP and LCP SH (or TH) intensities produced by a linearly polarized pump scales as the **fourth** (or **second**) power of the amplitude of the chiral resonance. The resonance amplitude is governed by two types of losses: radiative $\gamma_r$ and nonradiative $\gamma_{\mathrm{nr}}$. The radiative losses of a quasi-BIC resonance can be easily controlled by the structure perturbation strength (the notch size in our structures) that ensures the transition from a BIC state[21]. The nonradiative losses are associated with the material absorption and deviations of a real sample from an ideal periodic model. It has been highlighted that in case of negligible material absorption the nonradiative losses can be actively tuned with the sample size, which is crucial for observing resonances with extremely high Q-factor[30]. The inverse total Q-factor of a resonance can be divided into radiative and nonradiative parts:

$$1/Q = 1/Q_\mathrm{r} + 1/Q_\mathrm{nr},$$
$$Q_\mathrm{r} = Q_0 \alpha^{-2}, \ Q_\mathrm{nr} = \omega_\mathrm{r}/2\gamma_\mathrm{nr}, \tag{30}$$

where $\alpha$ is the asymmetry parameter controlled by the size of the notch and $\omega_r$ is the real part of the resonance frequency. According to the temporal coupled-mode theory[23,43], the resonance amplitude $a$ is proportional to:

$$a \propto \alpha Q |\mathbf{E}_\mathrm{inc}|, \tag{31}$$

and thus can be altered by changes in radiative and nonradiative losses. It is straightforward to derive that for the fixed radiative losses $\gamma_\mathrm{r} = $ const and $\alpha = $ const, we have:

$$a \propto (1 + \gamma_\mathrm{nr}/\gamma_\mathrm{cr})^{-1}, \tag{32}$$

where $\gamma_\mathrm{cr} = \omega_r/2Q_\mathrm{r}$ corresponds to the critical coupling regime.

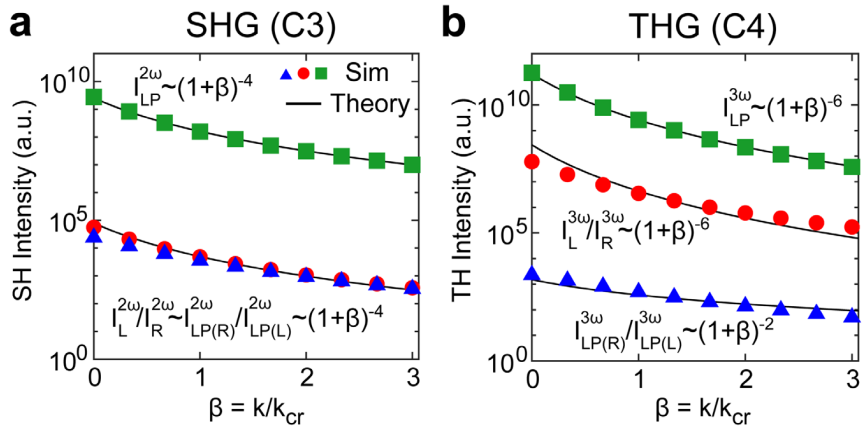

FIG. 5. **Scaling laws for nonlinear signal intensities in chiral metasurfaces. (a)** The dependence of $I_\mathrm{L}^{2\omega}/I_\mathrm{R}^{2\omega}$ (circles), $I_\mathrm{LP(R)}^{2\omega}/I_\mathrm{LP(L)}^{2\omega}$ (triangles) and $I_\mathrm{LP}^{2\omega}$ (squares) on the normalized extinction coefficient $\beta = k/k_\mathrm{cr}$ of the dielectric discs in the representative $C_3$ stacked metasurface with $H = 750$ nm, $\theta = 19^o$ with $k_\mathrm{cr} = 0.006$. Symbols denote the simulated results; lines show the theoretical scaling laws. **(b)** Same as **(a)** but for TH signal generated from the representative $C_4$-symmetric stacked metasurface with $H = 500$ nm, $\theta = 62^o$, and $k_\mathrm{cr} = 0.003$.

Dependencies of the absolute and relative nonlinear signal intensities on the extinction coefficient of the dielectric discs in the $C_3$ and $C_4$ metasurfaces are shown in FIG. 5, where the x-axis represents the normalized nonradiative losses $\beta \equiv k/k_\mathrm{cr} = \gamma_\mathrm{nr}/\gamma_\mathrm{cr}$. Note that at $k = k_\mathrm{cr}$ absorption at the frequency of the chiral mode reaches unity[26], and the Q-factor of the resonant mode is half smaller than that in the non-lossy case $k = 0$, thus $k_{cr}$ can be easily identified. Assuming that the resonant mode exhibits absolute chirality $a_\mathrm{R} = 0$ and $a_\mathrm{bg} = $ const, we substitute $a_\mathrm{L}$ in the form *(32)* into Eqs. *(1)*, *(3)*, (10), and *(14)* to compare the theory (lines) and simulation (symbols) results. As shown in FIG. 5a, the intensity of the SH signal generated under the LP pump $I_{LP}^{2\omega}$ scales as $(1 + \gamma_\mathrm{nr}/\gamma_\mathrm{cr})^{-4}$. The same is true for the ratio of the SH intensities generated under LCP and RCP pump $I_\mathrm{L}^{2\omega}/I_\mathrm{R}^{2\omega}$, as well as for the ratio of RCP and LCP SH signal intensities $I_\mathrm{LP(R)}^{2\omega}/I_\mathrm{LP(L)}^{2\omega}$ generated under the LP pump.

At the same time, the intensity of the TH signal generated under the LP pump $I_{LP}^{3\omega}$ scales similarly to the ratio of the TH intensities generated under the LCP and RCP pump $I_\mathrm{L}^{3\omega}/I_\mathrm{R}^{3\omega}$ in accordance with power law $(1 + \gamma_\mathrm{nr}/\gamma_\mathrm{cr})^{-6}$ FIG. 5b. However, the ratio of the RCP and LCP TH signal intensities generated under the LP pump $I^{3\omega}_{\text{LP(R)}}/I^{3\omega}_{\text{LP(L)}}$ matches a weaker trend $(1 + \gamma_{\text{nr}}/\gamma_{\text{cr}})^{-2}$. In addition to that, due to the high chirality of the resonant mode, one can estimate the nonlinear signal amplification associated with the resonant mode as the ratio of intensities under resonance-matching and resonance-mismatching polarization $I^{2\omega}_{\text{L}}/I^{2\omega}_{\text{R}}$, $I^{3\omega}_{\text{L}}/I^{3\omega}_{\text{R}}$. The simulation results are in good agreement with the theoretical predictions, a slight discrepancy occurs due to small deviations of the resonant modes from the state of perfect chirality (see SI-VI). These results prove that the high degree of circular polarization of the nonlinear signal under the LP pump in our structures is not an occasional feature and is determined by the amplitude of the chiral resonance. We conclude that SH generation supported by structures with $C_3$ rotational symmetry is the more promising effect that can be employed for LP to CP nonlinear signal conversion due to $I^{2\omega}_{\text{LP(R)}}/I^{2\omega}_{\text{LP(L)}} > I^{3\omega}_{\text{LP(R)}}/I^{3\omega}_{\text{LP(L)}}$, and hence higher DCP in the SH case.

**DISCUSSION**

Our work introduces and examines a novel category of highly efficient nonlinear all-dielectric metasurfaces. Combining rotational symmetry, strong quasi-BIC resonance, and chiral response, we create a metasurface that transforms a linearly polarized pump into a circularly polarized nonlinear signal. Absolute chirality of the degenerate quasi-BIC resonant modes is achieved through metasurface stacking. The chirality of the modes in stacked structures can be well predicted using the model of effective dipoles; furthermore, the resonant coupling model we present offers a comprehensive theoretical framework to explain the formation of chiral modes. Leveraging minimal inherent losses, this novel category of metasurfaces exhibits the potential for outperforming conventional nonlinear plasmonic counterparts. The proposed metasurface design obviates the need for a polarizer in optical setups to generate circularly polarized waves. By eliminating bulky components like polarizers, our approach sets the stage for further miniaturization of optical devices.

**MATERIALS AND METHODS**

All numerical simulations were conducted in the Electromagnetic Waves Frequency Domain module of COMSOL Multiphysics, 3D mode. In simulations, single unit cells of the metasurfaces with periodic boundary conditions were used with the adaptive spatial mesh controlled by the COMSOL. The mesh element size was manually set not to exceed 20 nm in the periodically patterned layers. The resonant modes were simulated using the COMSOL eigenfrequency solver. The resonant fields were integrated using COMSOL build-in integration nonlocal coupling functions to obtain the effective dipoles. All theoretical calculations and resonant dipoles post-processing for the resonance chirality evaluation were conducted in MATLAB. The normal incident plane wave scattering simulation was conducted using periodic ports with controlled excitation polarization (RCP, LCP, linear). To simulate the third and second harmonic generation, the fields obtained in the linear regime were used to evaluate the induced nonlinear polarization; this polarization was used as a signal source in a separate simulation at the nonlinear frequency. Scattering boundary conditions on top and bottom were applied in the nonlinear signal simulation. The amplitude of the far-field radiation propagating normally to the metasurface was calculated by taking the integral of the electric fields on the top and bottom boundaries of the simulation domain divided by the unit cell area. This allowed us to exclude the contribution of other diffraction channels.

**Data availability.** The main data supporting the findings of this study are available within the article and its Supplementary Information files. Extra data are available from the corresponding author upon reasonable request.

## ACKNOWLEDGMENTS

This work was supported by the Singapore University of Technology and Design for the Start-Up Research Grant SRG SMT 2021 169 and Kickstarter Initiative SKI 2021-02-14, SKI 2021-04-12; and National Research Foundation Singapore via Grant No. NRF2021-QEP2-02-P03, NRF2021-QEP2-03-P09, and NRF-CRP26-2021-0004. Dmitrii Gromyko acknowledges the support of SUTD-NUS Ph.D. RSS.

The authors thank Ilia Fradkin for insightful discussions and critical comments.

## AUTHORS' CONTRIBUTIONS

D.G., C.W.Q., and L.W. conceived the project idea. D.G. and J.F. developed the theoretical background. D.G. and J.S.L. performed numerical simulations. L.W. and C.W.Q. supervised the project. D.G., C.W.Q., and L.W. wrote the manuscript. All authors contributed to the interpretation of the results.

## CONFLICT OF INTEREST

The authors declare no conflicts of interest.